\documentclass[12pt,a4paper]{article}
\usepackage[utf8]{inputenc}
\usepackage{amsmath}
\usepackage{amsfonts}
\usepackage{amssymb}
\usepackage{makeidx}
\usepackage{graphicx}

\def\beq{\begin{equation}}
\def\eeq{\end{equation}}
\def\bea{\begin{eqnarray}}
\def\eea{\end{eqnarray}}

\begin{document}

\begin{center}
{\Large \bf General(ized) Hartman effect
  }

\vspace{1.3cm}

{\sf   Mohammad Hasan  \footnote{e-mail address: \ \ mhasan@isro.gov.in, \ \ mohammadhasan786@gmail.com}$^{,3}$,
 Bhabani Prasad Mandal \footnote{e-mail address:
\ \ bhabani.mandal@gmail.com, \ \ bhabani@bhu.ac.in  }}

\bigskip

{\em $^{1}$Indian Space Research Organisation,
Bangalore-560094, INDIA \\
$^{2,3}$Department of Physics,
Banaras Hindu University,
Varanasi-221005, INDIA. \\ }

\bigskip
\bigskip

\noindent {\bf Abstract}

\end{center}
In this letter we prove explicitly that if Hartman effect exists for  an arbitrary `unit cell' potential, then it also exist for a periodic system constructed using  the same `unit cell' potential repeatedly.  We further show that if Hartman effect exists, the tunneling time in the limiting case of a sufficiently  thick `unit cell' potential is same as that of its periodic system. This is true for any arbitrary value of the intervening gap between the consecutive `unit cell' of the periodic system. Thus generalized Hartman effect always occurs for a general potential constructed using multiple copies of single potential which shows Hartman effect.

\medskip
\vspace{1in}
\newpage

\section{Introduction}
The tunneling of a particle from a classically forbidden region is one of the most fundamental and earliest studied problem of quantum mechanics \cite {nordheim1928,gurney1928,condon,wigner_1955,david_bohm_1951}. However, how much time does a particle take to tunnel through a potential barrier is still and open problem. In the year 1962, Hartman studied the time taken by a wave packet to cross the classically forbidden region imposed by a  metal-insulator-metal sandwich \cite{hartman_paper} . He implied stationary phase (SPM) method to calculate the tunneling time.  It was found that for large thickness of the classically forbidden region, the tunneling time becomes independent of the thickness.  In an independent study,  this phenomena was also confirmed by Fletcher in later years  \cite{fletcher}. To present, this paradox is unresolved and is known as Hartman effect.  According to SPM, if the transmission coefficient through a potential barrier $V(x)$,  $0 \leq x \leq b$ is $t(E)=\vert t(E) \vert e^{i \delta (E)}$, then the tunneling time $\tau$ through the potential barrier is
\beq
\tau = \delta' + \frac{b}{2k}. 
\eeq  
Where $\delta' = \frac{d \delta}{dE}$ and wave vector $k= \sqrt{E}$  (we have chosen the unit $2m=1, \hbar=1, c=1$). For the case of a rectangular barrier $V(x)=V$ for $0 \leq x \leq b$, the tunneling time $\tau$ according to SPM yield \cite{dutta_roy_book}
\begin{equation}
\tau=  \frac{d}{dE} \tan^{-1} \left( \frac{k^{2}-q^{2}}{2kq} \tanh{q b}\right)
\end{equation} 
In the above equation, $q= \sqrt{(V-E)}$. We observe that $\tau \rightarrow 0$ as $b \rightarrow 0$ as expected. However,  when $b \rightarrow \infty$ we note
\begin{equation}
\lim_{b \rightarrow \infty} \tau = \frac{1}{qk},
\label{tt_qm}
\end{equation}
i.e. tunneling time is independent of the width of the barrier $b$ for a sufficiently opaque barrier. This is the  famous Hartman effect. 

Several studies have been conducted by different authors to understand this effect. The numerical monitoring of time evolution of the tunneling wave packet  have shown that the tunneling time agrees with the ones obtained by stationary phase method \cite{aquino_1998}.  In the case of two opaque barriers separated by a finite distance, it was found that the tunneling time is  independent of the separation of the barrier in the limit when the thickness of the barrier is large \cite{generalized_hartman}. This has given rise to the notion of generalized Hartman effect.  The phenomena of tunneling time being insensitive to  the intervening gap of the thick barriers for double or multi-barrier tunneling is known as generalized Hartman effect . For  multi-barrier tunneling  it has been shown that the total tunneling  time is  independent of barrier thickness and inter-barrier separation \cite{esposito_multi_barrier}. For the case of complex potential associate with elastic and inelastic channels, the Hartman effect has been found to occur for the case of weak absorption \cite{dutta}. For array of potentials associated with elastic and inelastic channels, the tunneling time saturates with the number of barriers \cite{our}.  

Several experiments have been conducted to test the finding of the theoretical results on tunneling time \cite{sl_prl,nimtz,ph,ragni,sattari,longhi1,olindo}. The tunneling time in all such experiments doesn't found to depend upon thickness of the tunneling region. Experiments were also conducted with double barrier photonic band gaps \cite{longhi2} and double barrier optical gratings \cite{longhi1}. The tunneling time was found to be independent of the gap between the two barriers and therefore also favors the generalized Hartman effect. For critical comments on generalized Hartman effect we refer to the articles  \cite{questions_ghf1,questions_ghf2,questions_ghf3}.   

The study of tunneling time from multiple barrier system \cite{generalized_hartman, esposito_multi_barrier} which has led to the concept of generalized Hartman effect are conducted by taking `unit cell' system as rectangular barrier which is known to  show Hartman effect. Our motivation for the present work arise to understand whether this is a general phenomena or not. In other words , if an arbitrary  `unit cell' system shows Hartman effect in the limit of increasing thickness,  then whether the periodic system constructed using multiple copies of 'unit cell'   will also show Hartman effect in the same limit. This leads to the question,  does the occurrence of Hartman effect from a potential also implies the occurrence of generalized Hartman effect. We found that this is always the case. In this letter we explicitly prove this using "uint cell' as barrier potential. Our result will be useful to study Hartman effect for many other complicated potentials.
 
We organize this letter as follows: In section \ref{ghf_section},  we establish the main result of this work  that the generalized Hartman effect always occurs for a potential that displays Hartman effect. In section \ref{result_section} we discuss the implications of this result. 

\section{Generalized Hartman effect}
\label{ghf_section}
In this section we prove that the generalized Hartman effect always exist for a potential that display Hartman effect. Consider that the transmission coefficient for an arbitrary potential $V(x)$ confined over the length $b$ is expressed as
\beq
t_{1} =\frac{1}{M_{1}}.
\eeq
Where $ M_{1}= \sqrt{v} e^{-i \delta}$ , $v \in R^{+} $  and $ \delta \in R$. $\frac{1}{v}$ is the transmission amplitude and $\delta$ is the phase of transmission coefficient.  If this `unit cell' potential $V(x)$ display Hartman effect, then the following quantity  will be independent of $b$ in the limit $b \rightarrow \infty$ 
\beq
\lim _{b \rightarrow \infty} \left( \delta' + \frac{b}{2k} \right) = \tau_{0}.
\eeq
$\tau_{0}$ is the tunneling time from the potential $V(x)$ in the limit $b \rightarrow \infty$. From the knowledge of $M_{1}$, the transmission coefficient for a periodic system made by $N$ repetitions of $V(x)$ is given by \cite{griffith_periodic}
\beq
t_{N} =\frac{e^{-ik N s}}{M_{1} e^{-ik s} U_{N-1}(\chi) -U_{N-2}(\chi)}.
\label{tn}
\eeq
Where,  
\beq
\chi =\sqrt{v} \cos{(\delta +k s)}.
\label{chi_eq} 
\eeq
In the above $s=b+L$ where $L$ is the intervening gap between the two consecutive `unit cell' potential of the periodic system.  From Eq. \ref{tn}, the phase of $t_{N}$ is given by
\beq
\Phi= \phi_{N} -k N s.
\eeq
Where,
\beq
\phi_{N} = \tan^{-1} \left [ \sqrt{v} \sin{(\delta + k s)} \frac{U_{N-1} (\chi)}{T_{N} (\chi)} \right ]. 
\label{phase_tn}
\eeq
The tunneling time according to SPM to traverse the length $(N-1)s+b$ of the periodic system is   
\beq
\tau_{N}= \left (\phi_{N}' - \frac{N s}{2k} \right) + \frac{(N-1)s+b}{2k} .
\label{tau_n1}
\eeq
In the above  $\phi_{N}' = \frac{d \phi_{N}}{dE}$ and the term in the parenthesis is the phase delay time.  Eq. \ref{tau_n1} is simplified to
\beq
\tau_{N}= \phi_{N}' - \frac{s}{2k}+ \frac{b}{2k} .
\label{tau_n2}
\eeq
To investigate the generalized Hartman effect, we calculate the limiting value of $\phi_{N}'$ when $b \rightarrow \infty$.  As $b$ and $E$ are two independent quantities related to the tunneling problem, we write
\beq
\lim_{b \to \infty } \phi_{N}' = \frac{d}{dE} \left ( \lim_{b \to \infty } \phi_{N} \right ) .
\label{phin_limit1}
\eeq
We note that for $b \rightarrow \infty$ the transmission from the potential $V(x)$ is expected to vanish i.e. $t_{1} \rightarrow 0 $. This implies $v \rightarrow \infty$ when $b \rightarrow \infty$. Thus for $b \rightarrow \infty$ we have $\vert \chi \vert \rightarrow \infty $ from Eq. \ref{chi_eq} provided $(\delta+ k s) \neq (2p+1) \frac{\pi}{2}$ , $p \in \{0, I^{+} \}$. For $b \rightarrow \infty$ , the condition of $(\delta+ k s)$ being odd integer multiple of $\frac{\pi}{2}$ never satisfies `for those potential that display Hartman effect'. Therefore by our initial assumption on the `unit cell' displaying Hartman effect,  we always have $ \lim_{b \rightarrow \infty} \vert \chi \vert \rightarrow \infty$. Therefore ,  
\beq
\lim_{b \to \infty } \frac{U_{N-1}(\chi)}{ T_{N} (\chi)} =  \lim_{\chi \to \infty } \frac{U_{N-1}(\chi)}{ T_{N} (\chi)} = \frac{1}{\chi}.
\label{chebyshev_limit}
\eeq
Using the above result in the expression of $\phi_{N}$ (Eq.\ref{phase_tn}) we get 
\beq
\lim_{b \to \infty } \phi_{N} = \delta + k s .
\label{blarge_phin}
\eeq
Plugging Eq. \ref{blarge_phin} in Eq. \ref{tau_n2} we arrive at,
\beq
\lim _{b \to \infty} \tau_{N}= \lim _{b \to \infty} \left (\delta' + \frac{b}{2k} \right ) = \tau_{0} .
\label{taun_limit}
\eeq
Eq. \ref{taun_limit} proves the  generalized Hartman effect for the periodic potential system. The tunneling from the periodic system is independent of $L$ and is equal to the tunneling time of the `unit cell' system in the limit $b \rightarrow \infty$. 

This is to be noted that the result represented by Eq. \ref{taun_limit} have been found  in \cite{generalized_hartman, esposito_multi_barrier} for real `unit cell' potential and in \cite{layered_pt, pt_hartman} for $PT$-symmetric `unit cell' potential. In all these cases, the `unit cell' potentials show Hartman effect. Thus our result that if a `unit cell' potential show Hartman effect, then its periodic system will also show Hartman effect have already found by different authors for specific cases of `unit cell' potential. Moreover, its quantitative representation by Eq. \ref{taun_limit} have also been noted by these authors.   In the present letter, we have shown this as a general result.   
\section {Results and Discussions}
\label{result_section}
We have given the proof that if a `unit cell' potential system displays Hartman effect , then the generalized Hartman effect always occur for the  periodic system made using  the multiple copies of the same `unit cell' potential repeatedly . We have given the proof  for an arbitrary $N$ repetitions of the `unit cell' system. This result has several implications in understanding the tunneling time by stationary phase method. As a special case we take the example of a rectangular barrier which is known to display Hartman effect. If the rectangular system is arranged as super periodic fashion (in which a periodic system repeats periodically and the new generated system further repeats periodically and this operation continue for arbitrary times \cite{spp}), the generated arrangement will show generalized Hartman effect. Similarly, the case of fractal potential such as Cantor-$\frac{1}{3}$, general Cantor , Smith Volterra Cantor (SVC) potential, general SVC potential etc. will  also show generalized Hartman effect. 

We have shown that for arbitrary $N$ repetitions of the `unit cell' potential, the tunneling time for the periodic potential is same as that of its `unit cell' potential when the `unit cell' potential is sufficiently thick. This quantitative result has simple qualitative interpretation. When the thickness of the `unit cell' potential is much larger as compared to the intervening gap between the two `unit cell' potential, the incident wave packet begins to see the periodic potential system as a single potential system. As the tunneling time from the `considered' single potential system (unit cell) is independent of  thickness for large thickness, the net tunneling time is the same as that of single potential system. Therefore the generalized Hartman effect is not a distinct feature of specific periodic system and can be considered as Hartman effect. However, why does a potential system display Hartman effect is paradoxical result and needs more investigations.

{\it \bf{Acknowledgements}}: \\
MH acknowledges supports from Director-SSPO  for the encouragement of research activities. BPM acknowledges the support from MATRIX project (Grant No. MTR/2018/000611), SERB, DST Govt. of India..

\end{document}